\documentclass[epj,nopacs]{svjour}
\usepackage{epsfig}
\usepackage{latexsym}
\usepackage[colorlinks=true,linktocpage=true,linkcolor=blue,citecolor=blue,allcolors=blue]{hyperref}
\usepackage{url}
\usepackage[utf8]{inputenc}
\usepackage{enumerate}
\usepackage{color}
\usepackage{xcolor}
\usepackage{microtype}

\usepackage{amsmath}
\usepackage{amssymb}

\usepackage[english]{babel}
\usepackage{url}

\begin{document}

 \title{
 A Chiral Mean-Field Equation-of-State in UrQMD: \\ Effects on the Heavy Ion Compression Stage
 }

 \titlerunning{CMF in UrQMD}

\author{Manjunath Omana Kuttan\inst{1,2,3} \and Anton Motornenko \inst{1} \and Jan~Steinheimer\inst{1} \and Horst~Stoecker\inst{1,2,4} \and Yasushi~Nara\inst{5} \and Marcus~Bleicher\inst{2,4,6}
}      

\institute{Frankfurt Institute for Advanced Studies, Ruth-Moufang-Str. 1, D-60438 Frankfurt am Main, Germany \and Institut f\"{u}r Theoretische Physik, Goethe Universit\"{a}t Frankfurt, Max-von-Laue-Str. 1, D-60438 Frankfurt am Main, Germany \and Xidian-FIAS international Joint Research Center, Giersch Science Center, D-60438 Frankfurt am Main, Germany \and GSI Helmholtzzentrum f\"ur Schwerionenforschung GmbH, Planckstr. 1, D-64291 Darmstadt, Germany \and Akita International University, Yuwa, Akita-city 010-1292, Japan \and Helmholtz Research Academy Hesse for FAIR (HFHF), GSI Helmholtzzentrum f\"ur Schwerionenforschung GmbH, Campus Frankfurt, Max-von-Laue-Str. 12, 60438 Frankfurt am Main, Germany}


\date{\today}

\abstract{
 It is shown that the initial compression in central heavy ion collisions at beam energies of $E_\mathrm{lab}=1-10A$~GeV depends dominantly on the underlying equation of state and only marginally on the model used for the dynamical description. To do so, a procedure to incorporate any equation of state in the UrQMD transport model is introduced. In particular we compare the baryon density, temperature and pressure evolution as well as produced entropy in a relativistic ideal hydrodynamics approach and the UrQMD transport model, where the same equation of state is used in both approaches. Not only is the compression similar if the same equation of state is used in either dynamical model, but it also strongly depends on the actual equation of state. These results indicate that the equation of state can be studied with observables which are sensitive to the initial compression phase and maximum compression achieved in heavy ion collisions at these beam energies.  
}

\maketitle

\section{Introduction}

 The theory of the strong interaction, Quantum Chromodynamics~(QCD), has a plethora of unknown properties and may offer a rich phase diagram that can only be revealed in the study of physical systems of very high baryon density and temperature. The details of the possible chiral and deconfinement QCD transitions are known, from first principle calculations, only in a narrow region close to vanishing baryon density~\cite{Borsanyi:2013bia,HotQCD:2014kol,Bazavov:2017dus} where both transitions appear as a smooth crossover. 
 
 To pin down the properties of hot and dense QCD matter has become the focus of various experimental programs ranging from laboratory experiments on earth to astrophysical data obtained from neutron star radii and binary neutron star mergers. Heavy ion collider programs at the BES at RHIC, NA61/Shine at CERN, CBM at GSI/FAIR, NICA in Russia and HIAF in China, and J-PARC-HI in Japan aim at studying the phase diagram of nearly isospin symmetric QCD matter at baryon densities of several times the nuclear matter ground state density and temperatures between 50 and more than 250 MeV.
 Here, the emphasis is mostly on the exploration of the phase transition of QCD matter from a 
confined hadronic phase to a phase where chiral symmetry is restored and quarks are eventually deconfined.

 At the same time, complementary astrophysical observations of binary neutron star mergers and supernova explosions can also create (iso-spin asymmetric) matter of comparable density and temperatures up to $50$ MeV \cite{ourletter}.
 Similarly, to the heavy ion collisions, cold neutron star matter is dynamically compressed and heated in binary neutron star mergers~(BNSM) which were experimentally detected by measuring gravitational waves~\cite{LIGOScientific:2017vwq,LIGOScientific:2020aai,Bauswein:2012ya,Most:2018eaw}.
 
 While the systems created in such different scenarios vary in size over many orders of magnitude, they share a common unknown and defining property, the equation of state (EoS) of dense nuclear matter. Extracting the equation of state and its properties, like phase transitions or softest points, has been a defining challenge to many experimental and theoretical programs over the last decade. 
 
 To do so, model simulations that can incorporate different possible equations of state are compared to experimental observables. The challenge in such an approach is that many features not related to the EoS, such as microscopic properties, unknown transport parameters or boundary conditions, are not well constrained. In the current 'state-of-the-art' of ultra-relativistic heavy ion collisions, the dynamical evolution of the collisions is divided into roughly three phases \cite{Paiva:1996nv,Hama:2004rr,Bass:2000ib,Hirano:2001eu,Kolb:2003dz,Hirano:2004en,Hirano:2012kj,Nonaka:2006yn,Petersen:2008dd,Werner:2010aa,Gale:2013da,Shen:2014vra}. In this so-called 'standard model of ultra-relativistic heavy ion collisions', the reaction starts with an initial non-equilibrium phase where the kinetic energy of the two incoming nuclei lose a fraction of their longitudinal momentum and create a pre-equilibrium fireball. This phase is usually described by string models or QCD inspired non-equilibrium approaches, e.g. via a color glass condensate model or quantum kinetic theory~\cite{Iancu:2003xm,Gelis:2010nm,Albacete:2014fwa,Drescher:2006pi,Drescher:2006ca,Schenke:2012wb,Rybczynski:2013yba,Moreland:2014oya}.
 Due to its violent non-equilibrium nature this phase of the reaction does generally not depend on the equation of state. After this energy deposition and a sufficient equilibration time, the near-equilibrium evolution can be described by (viscous) hydrodynamic or transport theoretical approaches. Here, an equation of state and transport properties of the medium can be included in the simulations. Finally once the system has hadronized, hadronic rescattering and the final freeze-out phase occurs \cite{Bass:2000ib,Steinheimer:2017vju}. As described above in this approach the EoS enters in the well defined equilibrium phase. Of course the applicable degrees of freedom vary with the collision energy.

 However, this 'standard picture' is only well justified at very high beam energies, i.e. when the initial interpenetration time of the incoming nuclei is very short and can be well separated from the subsequent expansion. This is generally found to be the case of heavy ion collisions above $\sqrt{s_\mathrm{NN}}\approx 15$ GeV (corresponding to ${\mathrm{E_{lab}} \gtrapprox 100 A \mathrm{GeV}}$), where mainly the energy of the incoming nuclei is stopped while the baryon number of the participant nuclei is observed far from mid-rapidity. At significantly lower beam energies, the interpenetration time can last as long or even longer than the expansion phase. In such a scenario a large amount of the baryon number is stopped in the central collision region and a system of high baryon density is created around mid-rapidity. 
 
The lower beam energies are exactly what is needed to study the EoS at the highest baryon densities. This also means that here the initial compression phase can not be separated from the expansion stage and the observables will therefore also be dependent on the equation of state in the initial compression phase. In particular this will be true in the presence of a phase transition. 
 
 It is therefore necessary to study the effects of the EoS on the initial compression at lower beam energies and also to devise new methods on how the dynamical evolution of such a system can be described. To achieve both of the above, a consistent treatment of the equation of state throughout the entire collision is necessary. 

 The challenges for the present paper are twofold: 
 \begin{itemize}
     \item First, a new method is introduced on how a realistic chiral mean field equation of state (CMF-EoS) can be incorporated in a non-equilibrium Quantum-Molecular-Dynamics transport approach (UrQMD).
     \item Second, the densities and temperatures achieved in this new approach (UrQMD-CMF) are compared with the evolution modeled with a relativistic (3+1) dimensional ideal fluid dynamical approach, where both approaches incorporate the same equation of state.
 \end{itemize}  
 Then we ask I) how sensitive is the initial compression on the equation of state to the different assumptions made in both approaches (i.e. full local thermalization in contrast to (non-)equilibrium transport dynamics) and II) up to which beam energy is a simple modeling of the systems evolution within a one-fluid hydrodynamic model equivalent to that of a full microscopic transport simulation.

\section{Methods}
In the following section the models used in the paper are described. These include the hydrodynamic model with its initialization routine, and the microscopic transport model UrQMD. Finally, the equations of state employed in these models are introduced alongside with the formalism for their consistent implementation in the dynamical models.

\subsection{Hydrodynamic approach}\label{ref:hydro}
The full 3+1D evolution of a heavy ion collision can be simulated by (ideal) relativistic hydrodynamics \footnote{Currently we neglect viscous and dissipative effects since we are mainly interested in the bulk evolution of the system}. These equations describe the conservation of energy and momentum given by

\begin{equation}
\partial_{\mu} T^{\mu \nu} =0\ ,
\end{equation}
as well as the conservation of the baryon four current
\begin{equation}\label{consr}
\partial_{\mu} j^{\mu}=0 \  .
\end{equation}

In the following, the \texttt{SHASTA} algorithm~\cite{Boris:1973tjt,Rischke:1995ir} is used for the flux-corrected relativistic numerical solution of the above equations.
The equations are solved on a Cartesian $200\times200\times200$ grid with a cell size ${0.2\times0.2\times0.2~{\rm fm^{-3}}}$ and the time-step is fixed to ${\delta t=0.4\times0.2=0.08}$ fm/c. To close this set of hydrodynamical equations an equation of state is necessary. The EoS can be treated as a free input to the equations, which is provided by a table, using only the constraints of strangeness neutral $n_S=0$ matter with a charge to baryon fraction of $n_Q/n_B=0.4$.

The hydrodynamic initial state is given by two counter streaming Lorenz-contracted Wood-Saxon distributions of baryon charge $n_{WS}$ (and corresponding energy density) representing the two colliding (cold) nuclei:
\begin{equation}
   n_{WS} = \gamma_{CM}  \frac{n_0}{1+\exp{\left(\frac{\Delta r-R}{a}\right)}}\,.
\end{equation}
Here, $\gamma_{CM}$ is defined by the collision energy in the center of mass ($CM$) frame. $\Delta r$ is the distance from the nucleus center and is Lorentz contracted along the z-axis. The parameters of the WS distribution are the nuclear saturation density $n_0$ as well as the nuclei radius $R$ and the surface thickness $a$. The values of $R=6.6$~fm and $a=0.5$~fm correspond to the known properties of $Au$ nuclei. This initialization procedure, contrary to the 'standard' hybrid model, allows for a hydrodynamic treatment, with inclusion of the EoS, of both the early entropy production and of the expansion stages, which is important for low collision energies where the interpenetration times are of the same magnitude as the systems lifetimes. 

 \subsection{Microscopic Transport approach}\label{ref:transport}
The non-equilibrium microscopic description of the heavy ion collisions is done with the \texttt{UrQMD} transport approach~\cite{Bass:1998ca,Bleicher:1999xi}. \texttt{UrQMD} is based on the covariant propagation of hadrons on classical trajectories in combination with stochastic binary scatterings, color string formation and resonance excitation and decays. The hadrons interaction criteria are based on a geometric interpretation of their scattering cross section. The cross sections for these scatterings are taken either from experimental measurements where available \cite{ParticleDataGroup:2020ssz}, or are calculated e.g. from the principle of detailed balance.
In its default setup the model corresponds to a hadronic cascade and can be readily used to describe the final state spectra of hadrons over a wide range of beam energies. It was shown that the effective equation of state of the \texttt{UrQMD} 'cascade model' corresponds to a Hadron Resonance Gas (HRG) with the respective degrees of freedom \cite{Bravina:2008ra}.

Extending the equations of motion to non-trivial hadronic interactions, and consequently to any possible equation of state, is not straightforward. 
Early, a non-relativistic QMD approach~\cite{Aichelin:1991xy} was developed to incorporate a density dependent Skyrme interaction~\cite{konopka}.
In this QMD part of the \texttt{UrQMD} model, the change of momenta of the baryons, due to a density dependent potential, is calculated using the non-relativistic equations of motion:

\begin{eqnarray}\label{motion}
\dot{\textbf{r}}_{i}=\frac{\partial  \langle H  \rangle}{\partial\textbf{p}_{i}},
\quad \dot{\textbf{p}}_{i}=-\frac{\partial  \langle H \rangle}{\partial \textbf{r}_{i}},
\end{eqnarray}

where {\it $\langle H \rangle$} is the total Hamiltonian function of the system. The Hamiltonian of each baryon, $H_i=E^{\mathrm{kin}}_i + V_i$, comprises the kinetic energy and the mean field potential energy $V_i=E_{\mathrm{field}}/A$ of the baryon $i$. The mean field potential energy per baryon can be related to a density dependent single particle energy:
\begin{equation}
    U_i(n_B)=\frac{\partial (n_B  \cdot V_i(n_B))}{\partial n_B}\,.
\end{equation}

In the Skyrme \texttt{UrQMD} approach \cite{Aichelin:1991xy,konopka,Bass:1998ca} the density dependence of the single particle energy for all baryons is given by a simple form:

\begin{equation}
\label{eq:usky}
    U_{\mathrm{Skyrme}}(n_B)= \alpha (n_B/n_0) + \beta (n_B/n_0)^{\gamma}\,.
\end{equation}

Two out of the three parameters ($\alpha$, $\beta$ and $\gamma$) are usually constraint by the nuclear matter saturation density and binding energy, while the remaining unconstrained property is the nuclear incompressibility, defining the so-called stiffness of the EoS.
Such a single free parameter approach to describe the equation of state of dense QCD matter has a significant shortcoming: the equation of state for densities above nuclear saturation is fixed by parameters which are defined solely at saturation density.
A similar problem occurs when a purely nuclear relativistic mean field model is implemented in QMD \cite{Nara:2019qfd,Nara:2020ztb}, although such an approach does also allow for the inclusion of additional degrees of freedom and thus a more complex phase structure. 
Recently, another idea has been put forward where additional terms are added in equation eq.(\ref{eq:usky}) which allow for describing non-trivial features like a phase transition in the potential \cite{Sorensen:2020ygf}. However, this density functional approach suffers from a serious problem characteristic to the Skyrme potential: the speed of sound of this EoS eventually becomes superluminal at large baryon densities, even at $T=0$. Below we introduce a different way to replace the limited Skyrme potential by a, more realistic, density dependent equation of state.

Once the mean field potential is known, the change of momentum of each baryon in accord with Hamiltons equations of motion can be calculated as:

\begin{eqnarray}
\dot{\textbf{p}}_{i}=-\frac{\partial  H }{\partial \textbf{r}_{i}} =  -\frac{ \partial{ V(n_B) }}{\partial n_B}\cdot\frac{\partial n_B({\bf r}_i)}{\partial \textbf{r}_{i}}\,.
\end{eqnarray}

Besides the derivative of the mean field potential energy, only the local density and its gradient is  required for each baryon.\footnote{Ignoring a possible momentum dependence of the potential.} This is calculated by assuming that each particle can be treated as a Gaussian wave packet~\cite{Aichelin:1991xy,Bass:1998ca}. With such an assumption, the local interaction baryon density $n_B(r_i)$ at location ${\bf r}_i$ of the $i$-th particle in the computational frame is:
\begin{equation}
    n_B(r_i) = \left(\frac{\alpha}{\pi}\right)^{3/2}\sum_{j,\,j\neq i} B_j \exp{\left(-\alpha({\bf r}_i-{\bf r}_j)^2\right)} \, ,
\end{equation}

where $\alpha=\frac{1}{2 L^2}$, with $L=\sqrt{2}$ fm, is the effective range of the interaction. The summation runs over all baryons, $B_j$ is the baryon charge of the $j$-th baryon.

In the following, the QMD implementation will assume, for simplicity, that the mean field potential for all baryon types is the same as for the nucleons.

 \subsection{Equations of state}
 
 The present paper aims at estimating the role of a realistic and cosnsitent equation of state for the compression in heavy ion collisions. The two approaches to simulating heavy ion dynamics, introduced in sections \ref{ref:hydro} and \ref{ref:transport}, allow to incorporate the interactions via an EoS. In the following we concentrate on two EoSs, a simple ideal HRG, and a realistic Chiral Mean Field~(CMF) model, which incorporates all interactions essential for a realistic description of nuclear matter, neutron stars, and hot QCD matter.
 
 {\it The HRG model.} The Hadron Resonance Gas model is an approximation to confined hadronic QCD matter~\cite{Dashen:1969ep}. It is based on the assumption that a gas of interacting hadrons 
can be described (if the width of the resonances is smaller than the temperature) by inclusion of all hadron species and their resonances as explicit degrees of freedom in the partition function. This partition function then mimics the basic thermodynamic properties of QCD at low temperatures and small densities. The HRG was shown to successfully describe the properties of lattice QCD thermodynamics below the chiral transition \cite{Karsch:2003zq,Huovinen:2009yb,Borsanyi:2010bp,Ratti:2010kj,Alba:2015iva,Bazavov:2012jq,Huovinen:2017ogf,Vovchenko:2014pka}. However, due to a lack of many-body and long-range interactions the model is not able to describe basic features of QCD phenomenology such as a bound nuclear ground state or deconfinement. 
Multiple extensions of the model have been developed over the years (for a survey, see~\cite{Walecka:1974qa,Dutra:2012mb,Dutra:2014qga,Vovchenko:2015vxa,Vovchenko:2016rkn,Vovchenko:2017cbu,Poberezhnyuk:2017yhx}). However, all modifications have similar shortcomings as the few-parameter description of the EoS in the Skyrme model.
 As discussed earlier the \texttt{UrQMD} model in the cascade mode will have an equilibrium state that is equivalent to the HRG model description of QCD matter~\cite{Bravina:2008ra}. As in the cascade mode of \texttt{UrQMD} only elastic scatterings and resonance excitations occur\footnote{Here string formation is omitted.}, the HRG is a good approximation for the effective EoS of the model.
A comparison of the hydro simulations with a HRG equation of state and \texttt{UrQMD} in cascade mode has been used as a reference to study the effects of instant equilibration on the dynamics \cite{Petersen:2008dd}. Note,  it is the cascade mode that is commonly used to calculate the initial compression phase in the prevalent hybrid models of heavy ion collisions \cite{Steinheimer:2007iy}.
 
 {\it CMF model.} The Chiral Mean Field model~\cite{Papazoglou:1998vr,Steinheimer:2010ib,Motornenko:2019arp} is an approach for the description of QCD thermodynamics for a wide range of temperatures and densities. The effective degrees of freedom of the CMF model include a complete list of known hadrons as well as the three light quark flavors plus a gluon contribution. The CMF contains the transition between quarks and hadronic degrees of freedom, the liquid vapor transition in nuclear matter, as well as chiral symmetry restoration are driven by mean fields. Parity doubling introduces heavy parity partners to the baryons of the lowest octet~\cite{Steinheimer:2011ea,Aarts:2018glk}. The baryons and their parity partners interact via mesonic mean fields (attractive scalar $\sigma,~\zeta$ and repulsive $\omega,~\rho,~\phi$ meson exchanges). The effective masses of the parity partners depend on the chiral fields, therefore the partners become mass-degenerate as the chiral symmetry is restored. A detailed description of the CMF model with its parameters can be found in \cite{Motornenko:2020yme}.

 The CMF model describes many aspects of QCD phenomenology. It has been successfully applied in an analysis of lattice QCD data~\cite{Motornenko:2020yme}, the description of cold neutron stars~\cite{Motornenko:2019arp}, and has been employed as EoS in the hydrodynamic simulations of both heavy ion collisions and binary neutron star mergers~\cite{Seck:2020qbx,ourletter}.
 
 The effective masses of the ground-state  octet baryons and their parity partners (assuming isospin symmetry) read~\cite{Steinheimer:2011ea}:
 \begin{eqnarray}
  m^*_{b\pm} &=& \sqrt{ \left[ (g^{(1)}_{\sigma b} \sigma + g^{(1)}_{\zeta b}  \zeta )^2 + (m_0+n_s m_s)^2 \right]} \nonumber \\ 
  & \pm & g^{(2)}_{\sigma b} \sigma \ ,
  \label{eq:mass}
  \end{eqnarray}
  where the various coupling constants $g^{(*)}_{*b}$ are determined by vacuum masses and by nuclear matter properties. $m_0$ refers to a bare mass term of the baryons which is not generated by the breaking of chiral symmetry, and $n_s m_s$ is the ${\rm SU}(3)_f$-breaking mass term that generates an explicit mass corresponding to the strangeness $n_s$ of the baryon. The single-particle energy of the baryons, therefore, becomes a function of their momentum $k$ and effective masses: $E^*=\sqrt{k^2+m_b^{* 2}}$.

 Similar to the effective mass $m_b*$ which is modified by the scalar interactions, the vector interactions lead to a modification of the effective chemical potentials for the baryons and their parity partners:

  \begin{equation}
      \mu^*_b=\mu_b-g_{\omega b} \omega-g_{\phi b} \phi-g_{\rho b} \rho\,.
      \label{eq:chem}
 \end{equation}

 Note that the couplings of nucleons and hyperons to the mean fields were fixed to reproduce nuclear binding energies $E_0/B\approx-15.2$~MeV as well as the asymmetry energy $S_0\approx31.9$~MeV, and incompressibility $K_0\approx267$~MeV.

The phase diagram of the CMF model includes three critical regions: the nuclear liquid-vapor phase transition, chiral symmetry restoration, and the transition to quark matter~\cite{Motornenko:2019arp}. The model predicts two first-order phase transitions. The first one is associated with the nuclear liquid-vapor phase transition at  $n_B\sim n_0$. The second one appears at about four times the normal nuclear density $4 n_0$ due to the chiral symmetry restoration. This chiral transition however shows only a small latent heat and the critical endpoint of this transition occurs already at $T_{\rm CP}\approx 17$ MeV.

 \begin{figure}[t]
   \includegraphics[width=0.5\textwidth]{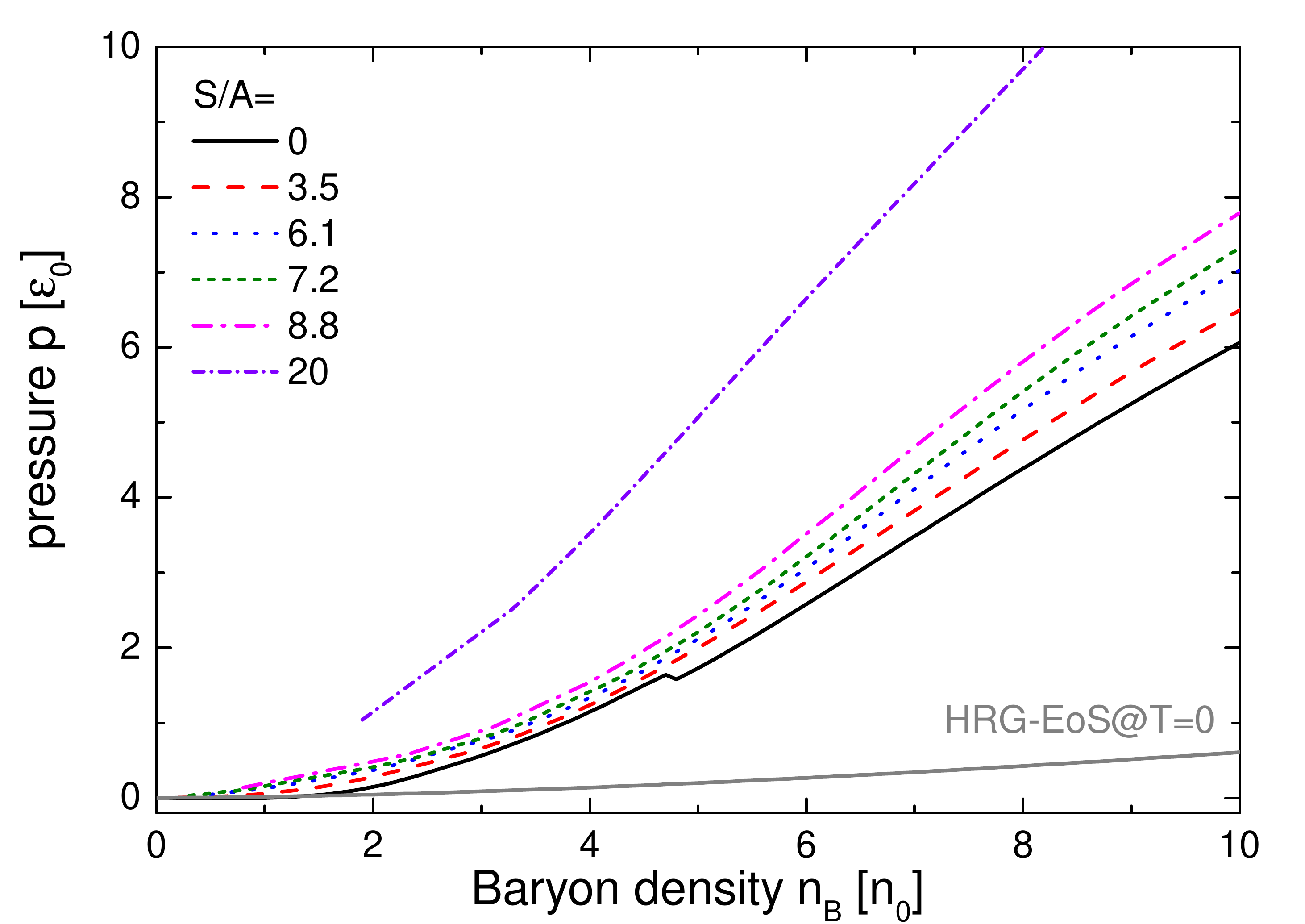}
   \caption{(Color online) The CMF equation of state, represented as pressure as a function of the baryon density, for different values of constant entropy per baryon (S/A). The black solid line corresponds to the CMF-EoS at zero temperature, where around 4$n_0$ a small kink in the pressure due to the phase transition can be observed. The same relation for the HRG-EoS at vanishing temperature is also shown as a grey line. The pressure in the HRG is substantially lower than in the CMF model.}
   \label{fig:eos_comp}
 \end{figure}

The CMF-EoS along different trajectories of fixed entropy per baryon is shown in Fig. \ref{fig:eos_comp}. This depiction is useful since one can see several relevant features in the CMF EoS. First, along the curve at zero entropy per baryon ($T=0$) a small kink in the pressure is observed which signals a very weak phase transition around four times saturation density. This kink disappears at higher entropies per baryon. Secondly for values of $S/A$ up to 10, the pressure only very mildly depends on the finite temperature and is dominated by the density dependence \footnote{Which supports our approach of assuming a mainly density dependent EoS in the implementation in UrQMD later.}. Finally, we also show the $T=0$ EoS in the HRG model as a grey line compared to the corresponding black line of the CMF. The CMF shows clearly a much larger pressure due to the mean field interactions, which will lead to observable effects in the dynamic simulations.

 In the CMF model the single nucleon potential is given by the interactions with the chiral and repulsive mean fields. At $T=0$, in the CMF model, it can be calculated from the self energy of the nucleons as:
 \begin{equation}
 \label{eq:cmfnp}
     \mathrm{U}_{CMF}= m_{N}^{*} - m_{N}^{vac} - \mu_{N}^{*} + \mu_{N} \,,
 \end{equation} 
 where $m_{N}^{vac}$ and $\mu_{N}$ are the vacuum mass and chemical potential of the nucleon calculated only from the charge constraints and $m_{N}^{*}$ and $\mu_{N}^{*}$ are the corresponding effective nucleon mass (eq.(\ref{eq:mass})) and effective chemical potential (eq.(\ref{eq:chem})) generated through the interactions with the scalar and vector mean fields.

\begin{figure}[t]
  \includegraphics[width=0.5\textwidth]{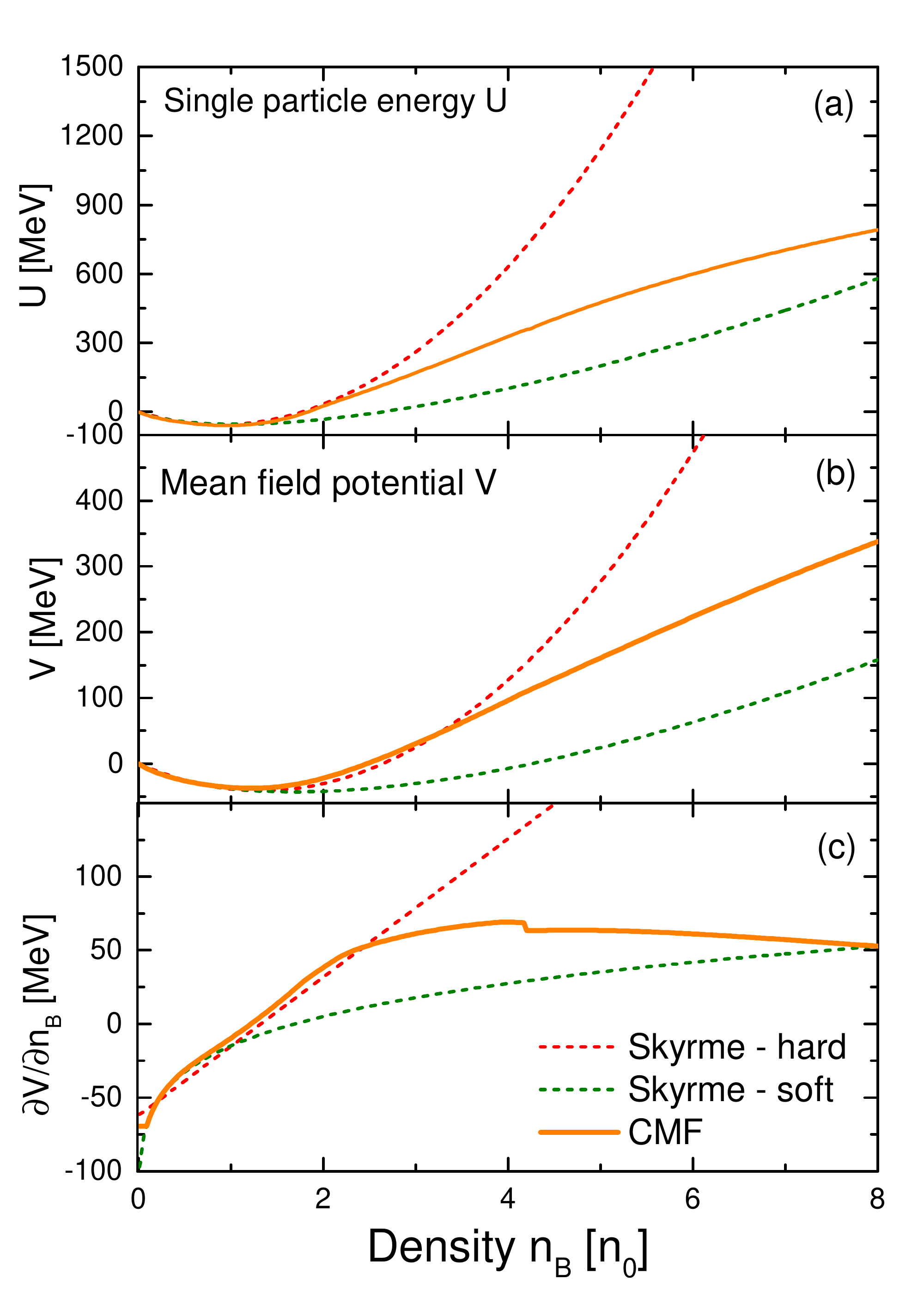}
  \caption{(Color online) (a) Nuclear potential $V$ as a function of baryon density $n_B$ in units of the ground-state baryon density for three different potentials (resulting in a different equations of state). 
  (b) The resulting field energy per baryon $E_{\mathrm{field}}/A$ for the three different equations of state.
  (c) Derivative of the field energy per nucleon with respect to the baryon density as a function of baryon density $\rho_B$ in units of the ground-state baryons density for three different potentials
  We show the Chiral mean field EoS (full orange line) in comparison to the well known hard Skyrme potential (dotted red line) and the soft Skyrme potential ( dotted green line).
  }
  \label{fig:skyrme-pot}
\end{figure}

 To set the stage, the CMF potential $\mathrm{U}_{CMF}$ is shown in panel (a) of Fig. \ref{fig:skyrme-pot} where we contrast the CMF single particle potential $\mathrm{U}_{CMF}$, as a function of baryon density $n_B$ in units of the ground-state baryon density, with two different Skyrme potentials $\mathrm{U}_{Sky}$ (resulting in different equations of state). We show the Chiral mean field EoS (full orange line) in comparison to the well known hard Skyrme potential (dotted red line) and the soft Skyrme potential (dotted green line).

\subsubsection{The CMF EoS in UrQMD}

\begin{figure*}[t]
  \centering
  \includegraphics[width=0.85\textwidth]{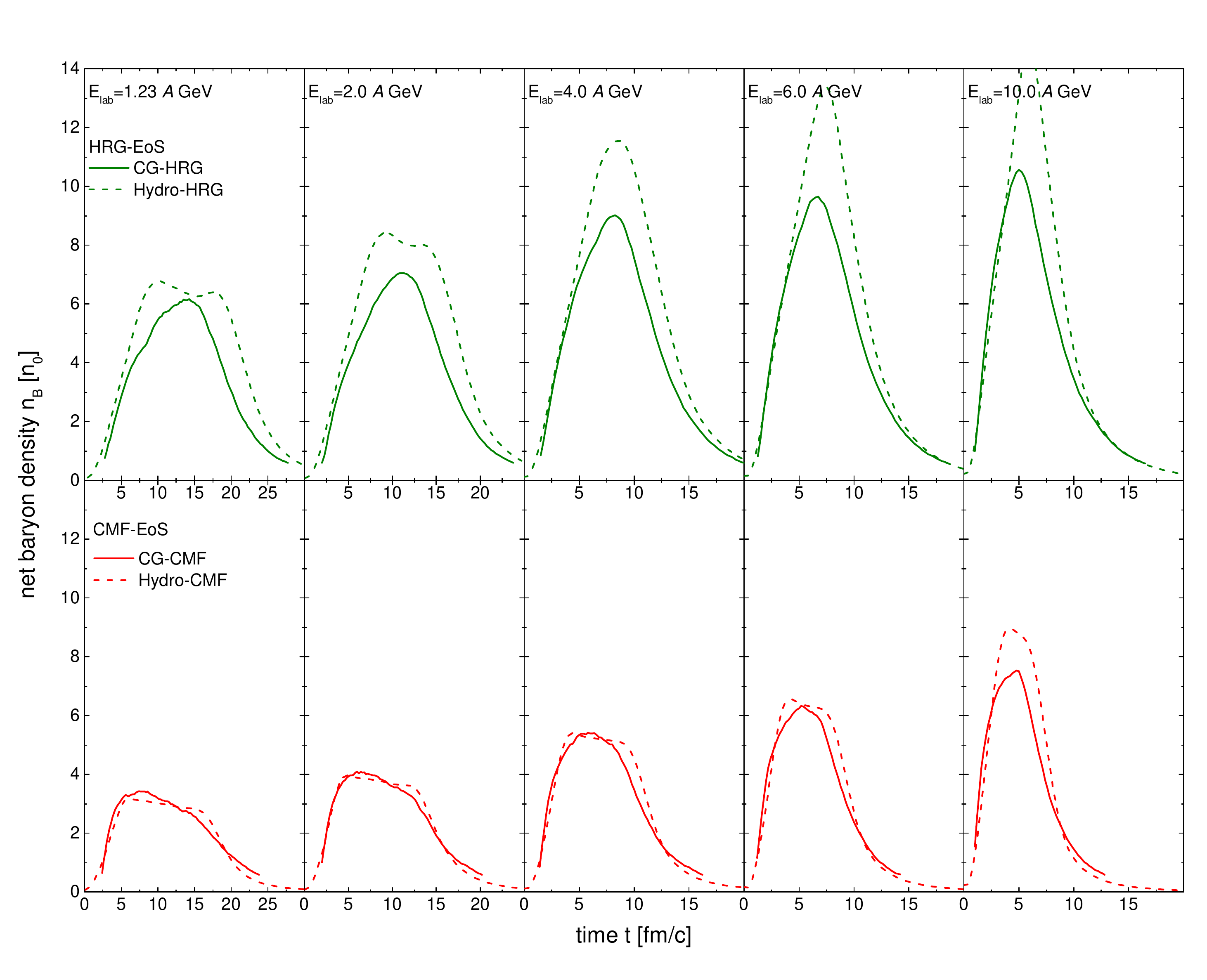}
\caption{(Color online) Time evolution of the baryon density in the central volume of the reaction for central Au+Au reactions at $E_\mathrm{lab}=1.23, 2.0, 4.0, 6.0, 10.0 A$ GeV (from left to right). The full lines show the results of coarse grained \texttt{UrQMD} simulations and the dashed line shows the results for one-fluid (3+1) dimensional hydrodynamic calculations for the same systems and energies. The green lines in the upper row are calculated using the hadron resonance gas EoS in hydro and for the conversion from $(\varepsilon,\rho_B)$ to the thermodynamic quantities, while the red lines in the lower row show the results for the CMF-EoS.  
  }
  \label{fig:rho-time}
\end{figure*}

To implement the CMF-EoS in the QMD part of the \texttt{UrQMD} model we essentially need to calculate the effective field energy per baryon of any particular model which can then be used in the QMD equations of motion. 
In the CMF model the nucleons interaction is described relativistically via scalar and vector mean fields which are not present in UrQMD. In addition, the CMF model is not only restricted to nucleons, thus, the single nucleon potential $\mathrm{U}_{CMF}$ as defined in eq. (\ref{eq:cmfnp}) is not suitable to calculate the relevant mean field potential that is required for the equations of motion.
Fortunately, the effective field energy per baryon $E_{\mathrm{field}}/A$ calculated from the CMF model can be used, i.e. the relevant quantity which enters the equations of motion:
\begin{equation}
    V_{CMF}=E_{\mathrm{field}}/A=E_{\mathrm{CMF}}/A - E_{\mathrm{FFG}}/A\,,
\end{equation}
where $E_{\mathrm{CMF}}/A$ is the total energy per baryon at $T=0$ from the CMF model and $E_{\mathrm{FFG}}/A$ is the energy per baryon from a free non-interacting Fermi-gas.
The resulting effective field energy per baryon, as a function of the baryon density, from the CMF model is shown as a solid line in panel (b) of figure \ref{fig:skyrme-pot}, again compared to the known curves from the hard and soft Skyrme EoS.
Finally, panel (c) of Fig. \ref{fig:skyrme-pot} shows the derivative of the field energy per nucleon with respect to the baryon density as a function of baryon density $n_B$ in units of the ground-state baryon density for the three different potentials.

\begin{figure*}[t]
  \centering
  \includegraphics[width=0.85\textwidth]{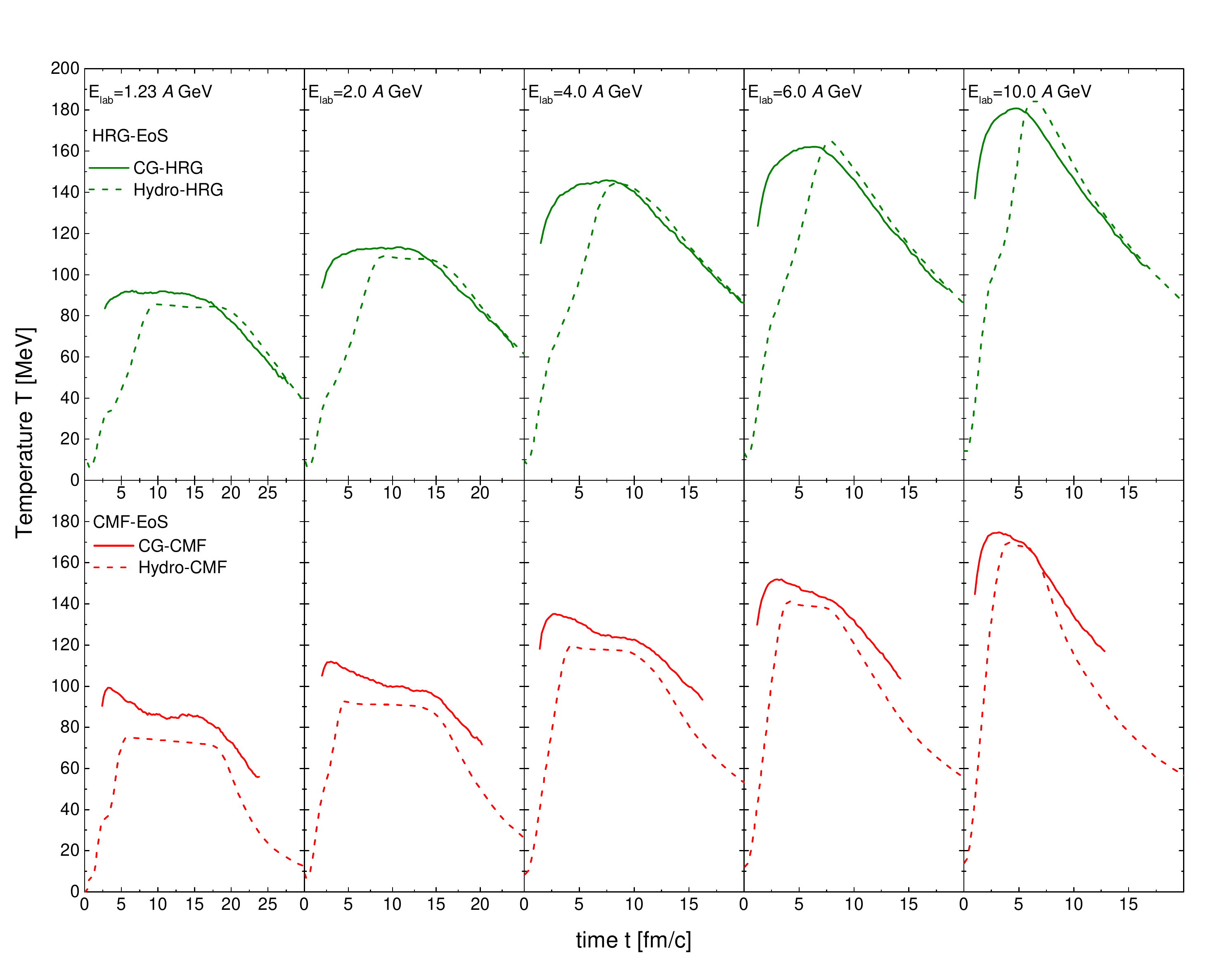}
\caption{(Color online) Time evolution of the average Temperature in the central volume of the reaction for central Au+Au reactions at $E_\mathrm{lab}=1.23, 2.0, 4.0, 6.0, 10.0 A$ GeV (from left to right). The full lines show the results of coarse grained \texttt{UrQMD} simulations and the dashed line shows the results for one-fluid (3+1) dimensional hydrodynamic calculations for the same systems and energies. The green lines in the upper row are calculated using the hadron resonance gas EoS in hydro and for the conversion from $(\varepsilon,\rho_B)$ to the thermodynamic quantities, while the red lines in the lower row show the results for the CMF EoS.  
In the coarse-graining procedure, only participants are used for the averaging, so the Temperature appears to 'jump' to a finite value.}
  \label{fig:T-time}
\end{figure*}

What can be observed is that the CMF-EoS shows a behavior similar to that of the soft Skyrme potential for sub-saturation (up to saturation) density, then becomes even stiffer than the hard Skyrme potential and finally shows a significant softening compared to the hard Skyrme which essentially becomes superluminal at large densities. Around four times nuclear saturation density the CMF-EoS shows a small kink in the derivative of the field energy per baryon which is due to the weak chiral phase transition at $T=0$. Since this transition is only very weak we expect no significant effects of this transition on the dynamic evolution, in particular at finite temperatures where the kink will be smeared out by the thermal energy.

Regarding its phase structure, the CMF model has several appealing features:
\begin{enumerate}
    \item A nuclear incompressibility compatible with experimental observations.
    \item A stiff super-saturation nuclear equation of state which is required to explain astrophysical observations.
    \item A "softening" of the equation of state at even higher densities due to the slow approach to the high density limit of a free gas of three quark flavors.
\end{enumerate}

Having now established a method in which any equation of state can be easily introduced in the QMD part of UrQMD, we will first study the dynamic evolution of bulk properties and their dependence on the EoS in the two dynamical approaches, hydrodynamics and microscopic transport.

\begin{figure*}[t]
  \centering
  \includegraphics[width=0.85\textwidth]{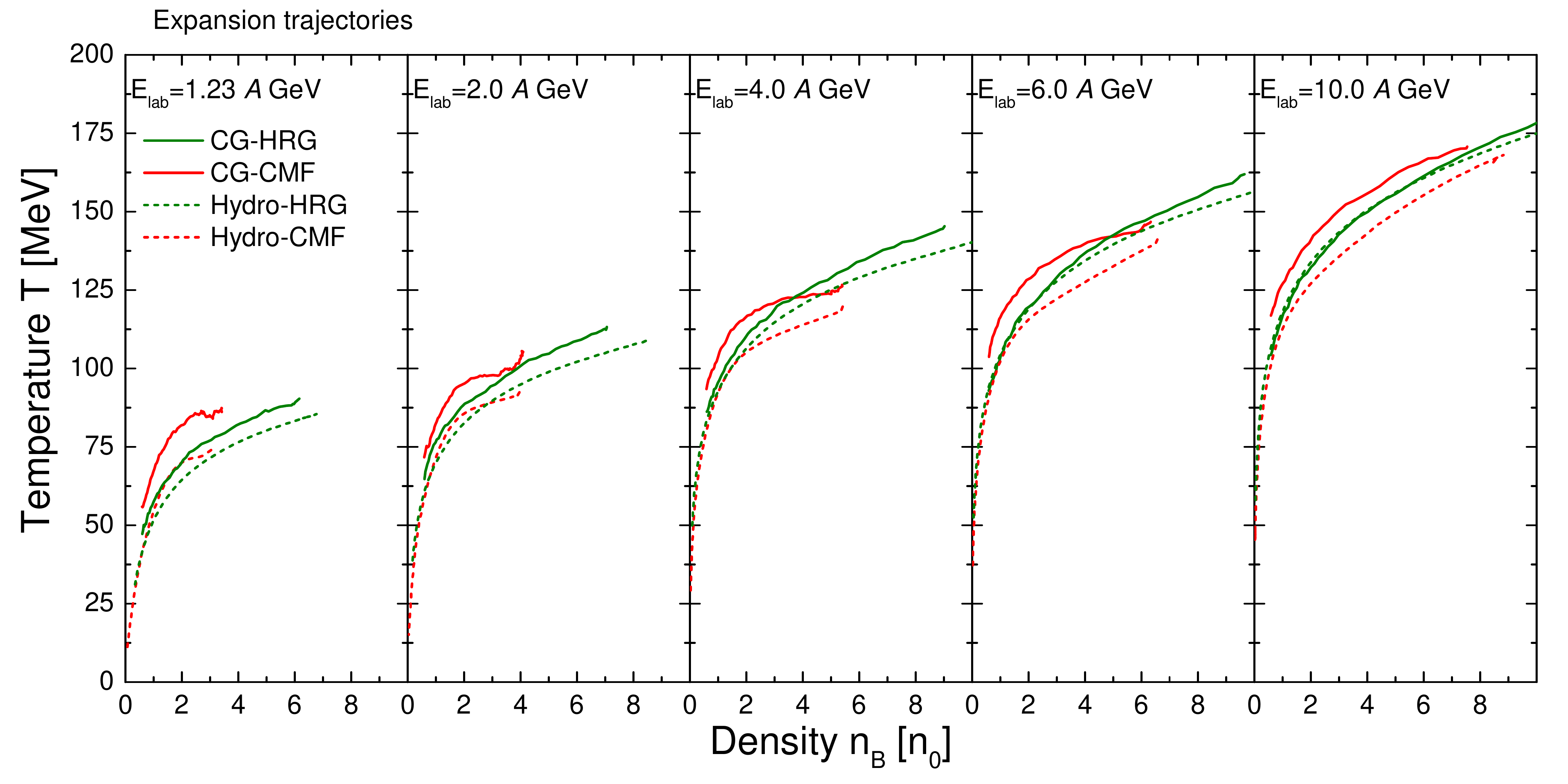}
  \caption{(Color online) Expansion trajectories (along the time evolution in the central cell) in the temperature-baryons density plane for central Au+Au reactions at $E_\mathrm{lab}=1.23, 2.0, 4.0, 6.0, 10.0 A$ GeV (from left to right). The full lines show the results of coarse grained \texttt{UrQMD} simulations and the dashed line shows the results for one-fluid (3+1) dimensional hydrodynamic calculations for the same systems and energies. The green lines show the results using the hadron resonance gas EoS for the conversion from $(\varepsilon,n_B)$ to the thermodynamic quantities, while the red lines show the results for the CMF-EoS.  
  }
  \label{fig:trho_isen}
\end{figure*}

\section{Results on the bulk evolution properties}
In this first work we will focus solely on bulk properties of the fireball created in central heavy ion collisions to establish the gross features of the CMF-EoS in the transport and hydrodynamic simulations. While it is known that other observables like radial flow and its higher moments can be very sensitive to the equation of state, the main focus of the present work is to establish the effect of the equation of state on the system created in different dynamical implementations of the same EoS. Studies of flow, cluster production as well as correlations and fluctuations will be explored in detail in future investigations.  

In the hydrodynamic model the expected properties of matter at different beam energies can be extracted in a straightforward way. Here, only one single event per beam energy, with impact parameter $b=2$ fm, is sufficient to average the thermodynamic properties in the central volume (a cubic volume of length $l=2$fm) of central Au+Au collisions at various beam energies. The local energy density and net baryon density are explicitly propagated in the hydrodynamic framework and quantities like the Temperature, pressure as well as entropy density can be directly and unambiguously related to these volume averaged densities via the equation of state.

In the microscopic transport treatment, the equivalent expectation values for the local energy and baryon number densities can also be calculated by a coarse graining procedure \cite{Endres:2015fna}. In this procedure, a large number of events of a given beam energy and centrality are calculated and the total energy and baryon density in the central volume, a cube of length 2~fm, can be calculated as sum of the energy and net baryon charge of participants in that volume. In this study, for a given beam energy, we use 1000 events with impact parameter less than 3.4 fm to perform the coarse graining. To extract the thermodynamic quantities like Temperature, pressure and entropy density a mapping to the effective equation of state, that is used, is necessary. In our simulations this mapping is done by using either the HRG-EoS (for the \texttt{UrQMD} cascade simulations) or the CMF-EoS (for the corresponding CMF-UrQMD simulation). Note that this procedure assumes that the system is close to local equilibrium which is not necessarily the case in the \texttt{UrQMD} transport model, especially at very early and late times. Thus the extracted values for the temperature, pressure and entropy density may (and as we later see will) vary due to deviations from equilibrium.

We begin in Fig. \ref{fig:rho-time} with a comparison of the time evolution of the baryon density in the central volume of the reaction for central Au+Au reactions at $E_\mathrm{lab}=1.23, 2.0, 4.0, 6.0, 10.0 A$ GeV (from left to right). The full lines show the results of the coarse grained \texttt{UrQMD} simulations, the dashed lines show the results for the one-fluid 3+1 dimensional hydrodynamic calculations for the same systems and energies. The green lines show the results using the HRG-EoS in hydro and for the coarse-graining conversion from $(\varepsilon,\rho_B)$ to the thermodynamic quantities, while the red lines show the results for the CMF-EoS.

\begin{figure*}[t]
  \centering
  \includegraphics[width=0.85\textwidth]{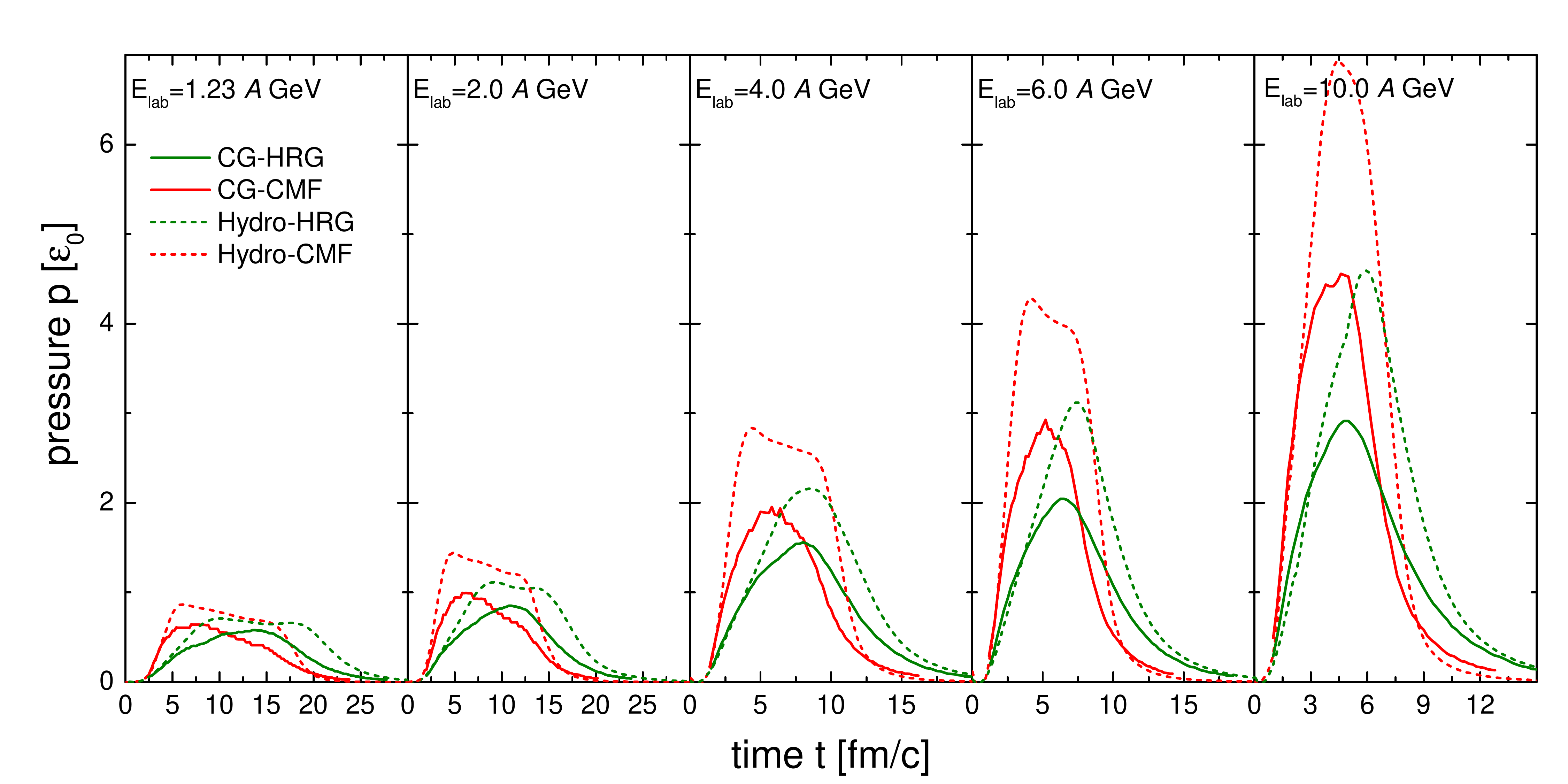}

  \caption{(Color online) Time evolution of the pressure in units of the ground-state energy density in the central cell of the reaction for central Au+Au reactions at $E_\mathrm{lab}=1.23, 2.0, 4.0, 6.0, 10.0 A$ GeV (from left to right). The full lines show the results of coarse grained \texttt{UrQMD} simulations and the dashed line shows the results for one-fluid (3+1) dimensional hydrodynamic calculations for the same systems and energies. The green lines show the results using the hadron resonance gas EoS for the conversion from $(\varepsilon,\rho_B)$ to the thermodynamic quantities, while the red lines show the results for the CMF-EoS.  
  }
  \label{fig:prho}
\end{figure*}

In the time evolution of the baryon density in figure \ref{fig:rho-time} one can clearly observe that the full hydrodynamic simulation and the transport simulation with the CMF-EoS give almost identical results up to the highest beam energies. Only for the beam energy of $E_\mathrm{lab}= 10.0 A$ GeV the transport simulation yields a smaller compression due to the effect of transparency which cannot be described in a 1-fluid simulation (although it is known that 3-fluid models can reproduce this effect). In the case of the HRG-EoS compared to the cascade model, the agreement of the density evolution is not as exact, yet still comparable. This is somewhat expected since the EoS in the cascade mode is not explicitly introduced and enters only implicitly by the set of degrees of freedoms. The fact that the transport model and the hydrodynamic model agree so well in the compression is not a trivial result but shows clearly, that the maximally reached compression, for low beam energies where transparency can be neglected, does to first order depend on the work that needs to be done against the pressure of the compressed system. The compression reached varies drastically, by almost a factor of 2, between the two equations of state used. This finding has important consequences since it means any observable that is sensitive to the maximal compression reached in these collisions would be a very good and almost model independent messenger for the equation of state of dense QCD matter.

Figure \ref{fig:T-time} shows the time evolution of the equilibrium temperature in the central volume of the same reactions. The colors and line styles are the same as in figure \ref{fig:rho-time}.
In the comparison of the (equilibrium) temperature, the differences between the hydro and transport approaches are more obvious. In particular at very early times, the temperature in the coarse grained transport simulations is significantly larger than in the fluid simulation. This can be understood as an effect of the non-equilibrium state of the microscopic transport which is then mapped on an equilibrium temperature. The effect of the non-equilibrium is not observed in the baryon density which is explicitly propagated and conserved in both approaches but in the temperature which is inferred implicitly. Nevertheless, after a few fm/c, even in the non-equilibrium approach the temperatures reached agree within 5-10 MeV. Notably the HRG simulations give also a systematically larger temperature, however the increase as compared to the CMF is only on the order of $~10$ MeV.

Having extracted the time evolution of the baryon density as well as the effective temperature, the expansion dynamics of the systems studied can also be depicted in the T-$n_B$ phase diagram.

Therefore, we explore the expansion trajectories along the time evolution in the central cell in Fig. \ref{fig:trho_isen} in the temperature-baryon density plane for central Au+Au reactions at $E_\mathrm{lab}=1.23, 2.0, 4.0, 6.0, 10.0 A$ GeV (from left to right). The full lines show the results of coarse grained \texttt{UrQMD} simulations and the dashed line shows the results for one-fluid (3+1) dimensional hydrodynamic calculations for the same systems and energies. The green lines show the results using the HRG-EoS for the conversation from $(\varepsilon,n_B)$ to the thermodynamic quantities, while the red lines show the results for the CMF-EoS. Note that for this comparison, we start the trajectories at the point of largest compression after which, in the case of the ideal hydrodynamics, they follow lines of constant entropy per baryon.

The most significant difference is that the HRG curves start at a much larger density. On the other hand, the trajectories become very close at lower densities. This means that at the time that the systems reach freeze out, at $n_B \leq n_0$, the thermodynamic conditions are very similar for the different models and equations of state. 

Much of the compression as well as consecutive expansion of the system strongly depends on the pressure reached during the initial phase. As we have seen a higher pressure in the EoS (harder EoS) leads to smaller densities. On the other hand the amount of radial as well as directed and elliptic flow produced will depend on the pressure which drives the expansion stage.
Finally, we show in Fig. \ref{fig:prho} the time evolution of the pressure in units of the ground-state energy density in the central cell of the reaction for central Au+Au reactions at $E_\mathrm{lab}=1.23, 2.0, 4.0, 6.0, 10.0 A$ GeV (from left to right). The full lines show the results of coarse grained \texttt{UrQMD} simulations and the dashed line show the results for one-fluid (3+1) dimensional hydrodynamic calculations for the same systems and energies. The green lines show the results using the HRG-EoS, while the red lines show the results for the CMF-EoS. Again, the extraction of the pressure from the local densities $(\varepsilon,n_B)$ is straightforward in the hydro model while for the coarse grained approach we assume local equilibrium and isotropic pressure which allows us to read of the effective pressure from the EoS table as described above.  

Most notably is that the maximum pressure is reached at different times, depending on the equation of state used. This is due to the maximum density also being reached at different times as shown in figure \ref{fig:rho-time}. This different time dependence of the pressure evolution is likely to have significant consequences on the generated flow, which we will study in detail in a forthcoming publication.

\subsection{Entropy production}

\begin{figure}[t]
  \centering
  \includegraphics[width=0.5\textwidth]{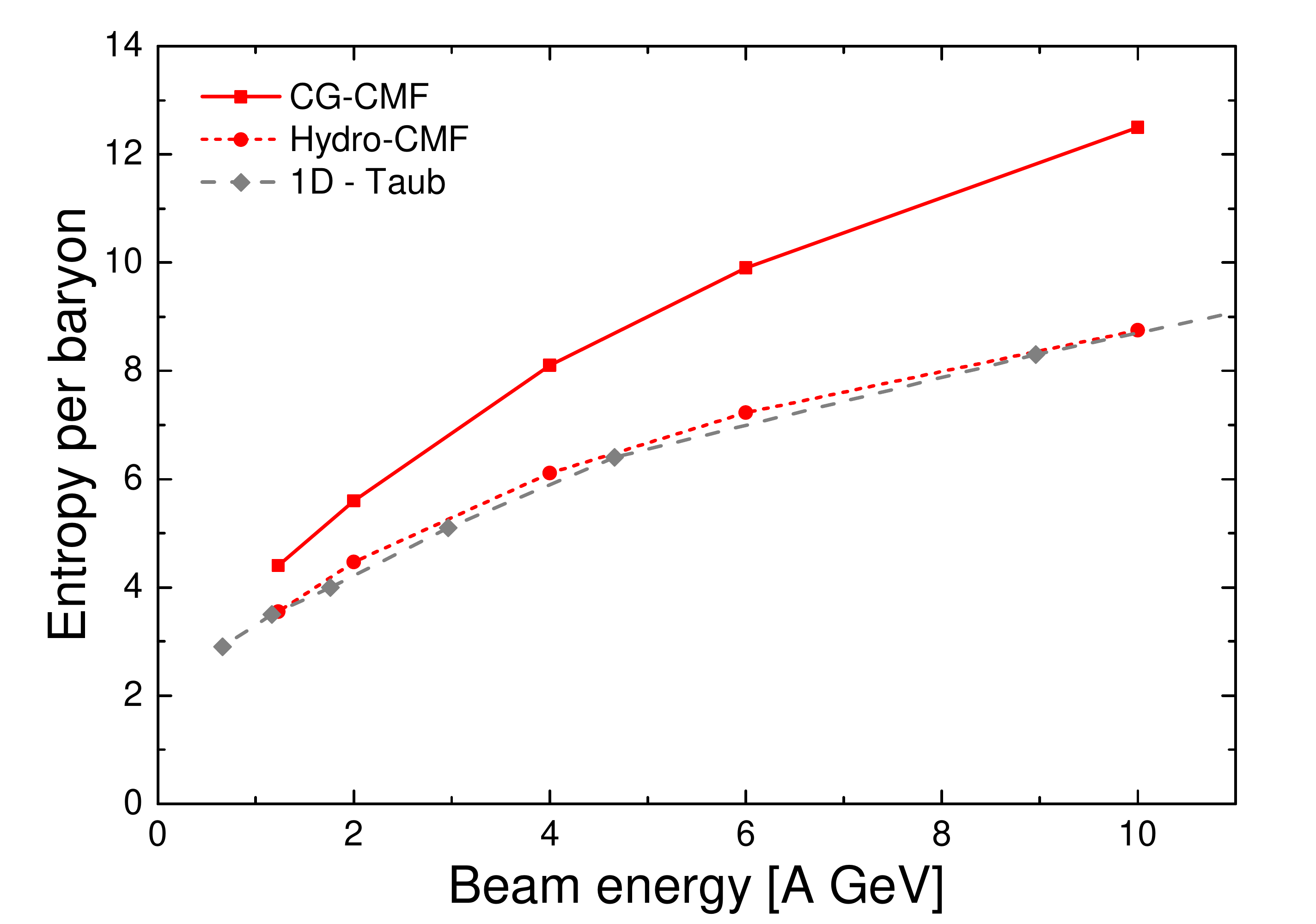}

  \caption{(Color online) Entropy production per baryon as a function of beam energy for central Au+Au reactions in the energy range from $E_\mathrm{lab}=1.23A$ GeV to $10 A$ GeV. The lines denote calculations using the coarse grained \texttt{UrQMD} model with CMF EoS (full red line), a one-fluid hydrodynamics calculation (dotted red line) and the one-dimensional shock model, i.e. the Taub adiabate (dashed grey line).
  }
  \label{fig:subera}
\end{figure}

As shown in figure \ref{fig:trho_isen} the expansion in both the hydro and transport models follows approximately the same isentropic trajectories. However, we expect that the final entropy per baryon will be different in the two approaches since the microscopic transport has a finite viscosity (shear and bulk) and the system will be only in partial chemical equilibrium at best. 
To complete the comparison, Fig. \ref{fig:subera} depicts the entropy production per baryon as a function of beam energy for central Au+Au reactions in the energy range from $E_\mathrm{lab}=1.23A$ GeV to $10 A$ GeV. The lines denote calculations using the coarse grained \texttt{UrQMD} model with CMF-EoS (full red line), the 3+1D one-fluid hydrodynamics calculation (dotted red line) and the one-dimensional relativistic shock model, i.e. the Taub adiabate (dashed grey line). For the hydrodynamics as well as the \texttt{UrQMD} coarse grained simulations, the entropy was extracted from the CMF model implicitly, as described for the temperature above, knowing the local energy and baryon densities. In the case of the hydro simulation, S/A as a function of time is essentially a constant throughout the expansion stage. In the transport simulation it only shows a slight increase. Here, we compare the values of S/A at the end of the expansion i.e. when the density drops below $n_B=n_0$.  
The full 3+1D ideal hydrodynamic simulation produces almost exactly the same entropy per baryon as the analytic 1-D shock solution (Taub adiabat). In general the entropy per baryon in the hydrodynamic case is smaller than in the non-equilibrium transport which is expected. The difference between these two scenarios grows with increasing energy which is also expected from the increasing transparency which leaves a smaller baryon number in the center of the collision zone. Furthermore, it is known that the system at late times can only be describes as being in partial chemical equilibrium. Mapping such a system onto an equilibrium EoS to calculate the entropy per baryon will yield larger values of the effective (equilibrium) S/A.

\section{Conclusion}

A method was introduced that enables us to implement any density dependent equation of state in the QMD part of the \texttt{UrQMD} model. It was shown that for low beam energies $E_\mathrm{lab} \leq 6A$ GeV, the bulk evolution of the density in this new description agrees very well with a relativistic 1-fluid simulation with the same equation of state. The effective temperature from the \texttt{UrQMD} simulation is slightly increased compared to the ideal hydrodynamic model due to non-equilibrium effects. 
Our results highlight the importance of the equation of state for the initial compression phase in nuclear collisions at low beam energies and, at the same time, provide a method on how it can be introduced in a consistent manner.

It was also shown that the expansion in both models follows closely an almost isentropic expansion as expected for the corresponding EoS. 

In the present study the CMF equation of state was used for both the hydro and \texttt{UrQMD} evolution. It describes properties of dense nuclear matter, astrophysical observables as well lattice QCD thermodynamics and includes a transition from hadronic to quark degrees of freedom.  

The total entropy per baryon produced in both scenarios was compared and it was found that the \texttt{UrQMD} model shows a slightly increased entropy production at low beam energies due to the non-equilibrium nature of the transport simulation. The entropy per baryon at the highest beam energies under investigation are significantly higher in the transport model due to the expected baryon transparency which transports the net baryon number away from mid-rapidity. However, the application of the presented approach is questionable for higher beam energies, where the stopping and energy deposition is dominated by partonic interactions (e.g. strings) for which the baryonic mean field QMD approach is not a suitable description.

Having established this new method enables us to now study the effect of different possible equations of state within the microscopic transport approach. The only assumptions are that the EoS is dominated by its density dependence i.e. fermions, and the effective potentials which govern the interactions have only a mild explicit temperature dependence. One should note that the density dependent forces that are assumed, are independent of the degree of equilibration reached throughout the collision.

In the future this method can readily be extended to include also a strong first order phase transition. This consistent description of the whole collision that does not require any ad-hoc matching of different phases will allow us to study possible observable signals of this transition in heavy ion collisions.

\begin{acknowledgement}

The authors thank Volker Koch for insightful discussions and
comments related to this work.
AM acknowledges the Stern-Gerlach Postdoctoral fellowship of the Stiftung
Polytechnische Gesellschaft. J.S. and MOK thanks the Samson AG and J.S. the BMBF through the ErUM-Data project for funding. MOK thanks GSI for funding. This work was supported by a PPP program of the DAAD.
MB acknowledges support by the EU-STRONG 2020 network.
Y.N. acknowledges support from JSPS KAKENHI Grant Number JP21K03577.
The computational resources for this project were provided by the Center for Scientific Computing of the GU Frankfurt and the Goethe-HLR.
\end{acknowledgement}

\end{document}